\documentstyle[12pt]{article}
\begin{document}

\title{\large On bound states in quantum field theory}
\author{G.V. Efimov\\
Bogoliubov Laboratory of Theoretical Physics,\\
JINR, 141980 Dubna, RUSSIA.}
\date{}
\maketitle
\begin{abstract}
The mechanism of formation of bound states in the relativistic
quantum field theory is demonstrated by the Yukawa field model.
It is shown that the weak coupling regime leads to the potential
picture, i.e. it is equivalent to the nonrelativistic limit in the
bound state problem. In the strong coupling regime the potential
picture is not valid and the method of bosonisation of fermion currents
(so-called $Z_2=0$ method) should be used. Essentially the nonlocal
fermion currents are found to be responsible for the origin of
bound states and ultraviolet convergence of fermion loops.
\end{abstract}

\newpage
\section{Introduction.}

The greatest efforts are currently being made in order to
describe physics of light and heavy bosons and baryons as
bound states of light and heavy quarks from the first
principles of QCD. The difficulty is that QCD like any
local quantum field theory is well defined as perturbation
expansion over an appropriate coupling constant, at the same time
bound states arise as a result of nonperturbative interactions.
Here I do not discuss the specific features of QCD such as confinement
of quarks and gluons, asymptotic freedom and behaviour at
large distances. The problem is more general: what is
the mechanism of formation of bound states in quantum field theory?
Great efforts were made to understand how bound states arise
in the formalism of quantum field theory and to work out effective methods
to calculate all characteristics of these bound states especially
their masses and quantum numbers. Unfortunately, we can establish
that there is no a well defined unique method, like
the Schr\"{o}dinger equation in the nonrelativistic quantum mechanics,
which can be used for this aim.

On the other hand, there are two well-known approaches to investigate
the bound state problem: the {\it Potential picture} (see, for example,
\cite{Lucha}) including Bethe-Salpeter equation (see, for example,
\cite{Greiner}) and the {\it Bosonization of fermion currents} or so-called
$Z_2=0$ approach (see, for example, \cite{Hayashi}). The potential
or quasipotential picture in the standard approach can be obtained by
taking into account the interaction of two particles-fermions
due to one-boson or one-gluon exchange.
This potential picture is formulated in the form of the Bethe-Salpeter
or quasipotential equation which turns into the Schr\"{o}dinger
equation in the nonrelativistic limit.

Let us explain in a few words the idea of the method of the {\it
Bosonization of fermion currents} or the $Z_2=0$ approach. Let a
quantum field system of fermions and bosons like quark-gluons,
electron-photon or nucleon-meson be given and the interaction be
described by the Yukawa-type Lagrangian. If a bound state
of a fermion-antifermion pair arises in one of these systems,
this bound state, being a bosonic state, should be
described by its own quantum bosonic field which is
absent in the initial Lagrangian. We can introduce this additional
quantum field into the initial Lagrangian adding the free Lagrangian of
this field and the interaction Lagrangian which is the product of
the boson field by the fermion current with appropriate quantum numbers.
The fact that this bosonic field is the bound state of the
fermion-antifermion pair means that the constant of renormalization
of the wave boson function is equal to zero, i.e. $Z_2=0$.

Effective practical application of this idea was always doubtful because
the renormalization constants are divergent in the local quantum field
theory where the fermion currents are usually chosen to be local.
Moreover, the relation between the potential picture
and the bosonization of fermion currents from the point of view
of quantum field theory is not clarified up to now.

Naturally, the task which should be solved in the first place is
to make clear the connection between these two approaches and to find
the conditions which separate them. The aim of this paper is to answer
this question by considering a relatively simple quantum field model.
This answer will help us to understand more deeply what kinds of
approximations are used in the standard well-known approaches.

An above-mentioned simple model under consideration is the Yukawa
interaction of the so-called "scalar" one-component fermions $\Psi$
interacting with scalar bosons $\phi$. The Lagrangian density is
\begin{eqnarray}
\label{int1}
L(x)&=&(\Psi^+(x)(\Box-M^2)\Psi(x))+
{1\over2}\phi(x)(\Box-m^2)\phi(x)\\
&+& g(\Psi^+(x)\Psi(x))\phi(x).\nonumber
\end{eqnarray}
The model makes it possible to retrace all details of arising of
bound states in quantum field theory. Generalization to the case
of the Dirac field has no difficulties of principle and leads to
technical problems connected with the algebra of $\gamma$-matrices only.
This model is superrenormalizable so that the renormalization procedure
has the simplest form. The main aim of this paper will be to understand
the general mechanism of arising of bound states in this quantum field
model and to outline possible methods to investigate this problem.

The model contains three dimensionless parameters:
\begin{eqnarray}
\label{int2}
&& \lambda={1\over4\pi}\left({g\over M}\right)^2,~~~~~~~~
\xi=\left({m\over M}\right)^2,~~~~~~~b=\left({\mu\over2M}\right)^2,
\end{eqnarray}
where $\mu$ is the mass of a supposed bound state. The parameter $\xi$
is supposed to be smaller then 1. Our aim will be to find the condition
under which the mass $\mu$ of the bound state belongs to  the interval
\begin{eqnarray}
\label{int3}
&& 0<\mu<2M~~~~~~~{\rm or}~~~~~~~0<b=\left({\mu\over2M}\right)^2<1,
\end{eqnarray}
i.e. this bound state should be stable.

The value of the coupling constant $\lambda$ plays
the crucial role to separate these two approaches: the potential
picture and the bosonization of fermion currents. Namely
\begin{itemize}
\item the {\it Potential picture} takes place for
\begin{eqnarray}
\label{int4}
&& \lambda={1\over4\pi}\left({g\over M}\right)^2\ll1
\end{eqnarray}
and $\mu=2M-\epsilon$ where
$$ \epsilon\ll 2M~~~~~~{\rm or}~~~~~~
1-b=\left[1-\left({\mu\over2M}\right)^2\right]\ll 1,$$
i.e., the mass excess is very small;
\item the {\it Bosonization of fermion currents} takes place for
\begin{eqnarray}
\label{int5}
&& \lambda={1\over4\pi}\left({g\over M}\right)^2\geq1
\end{eqnarray}
and the mass excess can be comparable with $2M$, i.e.
$$ 2M-\mu\sim M~~~~~~~~{\rm or}~~~~~~b\sim {1\over2}.$$
\end{itemize}

The result of this paper is that in the relativistic quantum field
theory the nonrelativistic potential picture takes place in the
weak coupling regime and the bosonization of fermion currents
takes place in the strong coupling regime.

In principle, this paper contains only the first rough scheme showing
the way how to calculate masses and quantum numbers of possible bound
states in quantum field systems. There are plenty of questions which
have to be answered.

This work is supported by the Russian Foundation for Fundamental
Research N 96-02-17435a.

\vspace{.5cm}

\section{The formulation of the problem}

For simplicity we consider the interaction of scalar fermions described
by the one-conponent field $\Psi(x)$ and bosons described by the field
$\phi(x)$. All consideration is given in the Euclidean metrics.
The total Lagrangian can be written in the form
\begin{eqnarray}
\label{for1}
 L[\Psi,\phi]&=&\int dxL(x)=
-(\Psi^+D^{-1}_M\Psi)-{1\over2}(\phi D^{-1}_m\phi)+g(\Psi^+\Psi)\phi,
\end{eqnarray}
$$ D^{-1}_M=D^{-1}_M(x-y)=(-\Box+M^2)\delta(x-y),$$
$$ D_M(x-y)=\int{dp\over(2\pi)^4}{e^{ip(x-y)}\over M^2+p^2}.$$

The object of our interest is the four-point Green function
\begin{eqnarray}
\label{for2}
&& G(x_1,x_2,y_1,y_2)\\
&& =\int D\Psi D\Psi^+\int D\phi~
\Psi^+(x_1)\Psi(x_2)\Psi^+(y_1)\Psi(y_2)~e^{L[\Psi,\phi]},\nonumber
\end{eqnarray}
where the normalization should be introduced
$$ D\Psi D\Psi^+\to{D\Psi D\Psi^+\over C},~~~~~~~
C=\int D\Psi D\Psi^+\int D\phi e^{L[\Psi,\phi]}.$$
The four-point Green function (\ref{for2}) contains all information
about possible bound states in the two-fermion channels. The particles
$\Psi$ can be called constituent particles. We shall be interested
in the bound states with definite quantum numbers which can be
defined as
\begin{eqnarray}
&& \Psi^+(x_1)\Psi(x_2)\to J_Q(x)=(\Psi^+V_Q\Psi)_x=
\Psi^+(x)V_Q(\stackrel{\leftrightarrow}{p}_x)\Psi(x),\nonumber\\
&& \stackrel{\leftrightarrow}{p}_x=
{1\over i}\left[\stackrel{\leftarrow}{\partial}_x-
\stackrel{\rightarrow}{\partial}_x\right],\nonumber
\end{eqnarray}
where the nonlocal vertex $V_Q(\stackrel{\leftrightarrow}{p}_x)$
defines the quantum numbers $Q$ of the current $J_Q=(\Psi^+V_Q\Psi)$.
It can be represented in the form
\begin{eqnarray}
\label{for3}
&& V_Q(\stackrel{\leftrightarrow}{p}_x)=
\int du\tilde{V}_Q(u)e^{iu\stackrel{\leftrightarrow}{p}_x},
\end{eqnarray}
The current $J_Q=(\Psi^+V_Q\Psi)$ can be represented as
$$ J_Q(x)=(\Psi^+V_Q\Psi)_x=(\Psi^+(x)V_Q(\stackrel{\leftrightarrow}
{p}_x)\Psi(x))=\int du~\Psi^+(x+u)\tilde{V}_Q(u)\Psi(x-u).$$

The Green function with quantum numbers $Q$ is defined as
\begin{eqnarray}
\label{for4}
&& G_Q(x-y)=\int D\Psi D\Psi^+\int D\phi~J_Q(x)J_Q(y)~
e^{L[\Psi,\phi]}.
\end{eqnarray}
This function has the following asymptotic behaviour
$$ G_Q(x-y)\sim e^{-\mu_Q\vert x-y\vert}~~~~~~{\rm for}~~~~~~
\vert x-y\vert\to\infty,$$
The mass of the state $J_Q=(\Psi^+V_Q\Psi)$ can be found as
\begin{eqnarray}
\label{for5}
&& \mu_Q=-\lim_{x\to\infty}{1\over\vert x\vert}\ln G_Q(x).
\label{mass}
\end{eqnarray}
The problem is to calculate the functional integral in the
representation (\ref{for4}) and calculate the mass $\mu_Q$ according
to (\ref{for5}).

If we consider perturbation expansion over the coupling constant $g$
for the four point Green function $G(x_1,x_2,y_1,y_2)$, we will
get a series of the Feynman diagrams describing an interaction
of two fermion particles $\Psi$. This series can be written in
the form of the Bethe-Salpeter equation \cite{Greiner}.
Bound states of two fermion particles in a channel
$J_Q(x)=(\Psi^+V_Q\Psi)_x$
can be found as solutions of this equation. Really, the Bethe-Salpeter
equation can effectively be investigated only in the one-boson
exchange approximation, i.e. in the weak coupling regime.

We proceed in another way, we shall consider
the generating functional (\ref{for4}).
Fortunately, it is possible to do the first integration over either
the fermion field $\Psi(x)$ or the boson field $\phi(x)$. Thus, we
can get two representations which are the starting points of two
approaches: the {\it Potential picture} and the {\it Bosonization of
fermion currents}.

\vspace{.5cm}
{\bf I. Potential picture}
\vspace{.5cm}

The integration in (\ref{for2}) over the fermion field $\Psi$ gives
\begin{eqnarray}
\label{for6}
&& \int D\Psi D\Psi^+~\Psi^+(x_1)\Psi(x_2)\Psi^+(y_1)\Psi(y_2)~\nonumber\\
&&\cdot\exp\left\{\int dx\Psi^+(x)\left(\Box-M^2+g\phi(x)\right)
\Psi(x)\right\}\nonumber\\
&& =\left[-S^+(x_1,y_2\vert\phi)S(x_2,y_1\vert\phi)
+S(x_2,x_1\vert\phi)S(y_2,y_1\vert\phi)\right]~\nonumber\\
&&\cdot\exp\left\{{\rm tr}\ln[1-g\phi D_M]\right\}.
\end{eqnarray}
The Green function $S(x,y\vert\phi)$ satisfies the equation
\begin{eqnarray}
\label{for7}
[\Box-M^2+g\phi(x)]S(x,y\vert\phi)=\delta(x-y)
\end{eqnarray}
with
$$ S^+(x,y\vert\phi)=S(x,y\vert\phi).$$
The Green function (\ref{for4}) can be written
\begin{eqnarray}
\label{for8}
G_Q(x-y)&=&-G_Q^{(P)}(x-y)+G_Q^{(A)}(x-y),
\end{eqnarray}
\begin{eqnarray}
&& G_Q^{(P)}(x-y)=\int du\int dv\int D\phi e^{L_P[\phi]}~\nonumber\\
&& \cdot\left\{\tilde{V}_Q(u)S(x+u,y+v\vert\phi)
\tilde{V}_Q(v)S(y-v,x-u\vert\phi)\right\},\nonumber
\end{eqnarray}
\begin{eqnarray}
&& G_Q^{(A)}(x-y)=\int du\int dv\int D\phi e^{L_P[\phi]}~\nonumber\\
&& \cdot\left\{\tilde{V}_Q(u)S(x+u,x-u\vert\phi)\right\}\cdot
\left\{\tilde{V}_Q(u)S(y+v,y-v\vert\phi)\right\}\nonumber
\end{eqnarray}
where
$$ L_P[\phi]=-{1\over2}(\phi D^{-1}_m\phi)+{\rm tr}\ln[1-g\phi D_M].$$
Here the functions $G_Q^{(P)}(x)$ and $G_Q^{(A)}(x)$ are said to be
"potential" and "annihilation" Green functions, respectively.
The approach based on the representation (\ref{for8}) shall be
called the {\it Potential picture}.

\vspace{.5cm}
{\bf II. Bosonization of fermion currents}
\vspace{.5cm}

The integration in (\ref{for4}) over the scalar field $\phi(x)$ gives
\begin{eqnarray}
\label{for9}
G_Q(x-y)
&=&\int D\Psi D\Psi^+e^{L_B[\Psi]}~(\Psi^+V_Q\Psi)_x(\Psi^+V_Q\Psi)_y,\\
L_B[\Psi]&=&-(\Psi^+D^{-1}_M\Psi)+{g^2\over2}(\Psi^+\Psi D_m\Psi^+\Psi).
\nonumber
\end{eqnarray}
where
\begin{eqnarray}
\label{for10}
&& (\Psi^+\Psi D_m\Psi^+\Psi)=\int dx\int dy
\Psi^+(x)\Psi(x)D_m(x-y)\Psi^+(y)\Psi(y)~,
\end{eqnarray}
and the following normalization should be introduced
$$ D\Psi D\Psi^+\to{D\Psi D\Psi^+\over C},~~~~~~~
C=\int D\Psi D\Psi^+e^{L_B[\Psi]}.$$
The approach based on the representation (\ref{for8}) shall be called
the {\it Bosonisation of fermion currents}.

\section{The Potential picture.}

The starting point of the Potential picture is the representation
(\ref{for8}).

\subsection{The Green function $S(x,y|\phi)$}

The first step is that the fermion loops should be neglected, so that
the Green functions $G_P(x-y)$ and $G_Q(x-y)$ are represented
\begin{eqnarray}
\label{pot1}
G_P(x-y) &=& \int d\sigma_{uv}[\phi]~\tilde{V}(u)
S(x+u,y+v\vert\phi)\tilde{V}(v)S(x-u,y-v\vert\phi),\nonumber\\
G_A(x-y) &=& \int d\sigma_{uv}[\phi]~\tilde{V}(u)
S(x+u,x-u\vert\phi)\cdot\tilde{V}(v)S(y+v,y-v\vert\phi)
\nonumber\\
d\sigma_{uv}[\phi] &=& dudvD\phi~e^{-{1\over2}(\phi(x)D^{-1}_m\phi)}.
\end{eqnarray}
We would like to stress now that the neglect of fermion loops
presupposes the dimensionless coupling constant $\lambda$ (\ref{int2})
to be small enough.

The solution of (\ref{for7}) can be represented in the form of the
following functional integral (see Appendix and, for example,
\cite{Dineykhan}):
\begin{eqnarray}
\label{pot2}
&& S(x,y|\phi)={1\over\Box-M^2+g\phi(x)}\cdot\delta(x-y)\\
&&=\int\limits_0^\infty {d\alpha\over8\pi^2\alpha^2}e^{-{\alpha\over2}M^2}
\int D\xi\exp\left\{-\int\limits_0^\alpha d\tau{\dot{\xi}^2(\tau)\over2}+
{g\over2}\int\limits_0^\alpha d\tau\phi\left(\xi(\tau)\right)\right\},
\nonumber
\end{eqnarray}
with the boundary conditions $\xi(0)=y,~~\xi(\alpha)=x$ and the
normalization
$$\int D\xi\exp\left\{-\int\limits_0^\alpha d\tau{\dot{\xi}^2(\tau)\over2}
\right\}=1.$$

\subsection{The Green function $G_P(x)$}

The function $G_P(x)$ after integration over $\phi$ has the form
$$ G_P(x)=\left({M\over8\pi^2x}\right)^2
\int d\Sigma_1 d\Sigma_2\exp\{W_{11}+2W_{12}+W_{22}\}\,$$
$$ W_{ij}={g^2\over8}\int\limits_{0}^{\alpha_i}d\tau_1
\int\limits_{0}^{\alpha_j}d\tau_2
D_m\left(\xi_i(\tau_1)-\xi_j(\tau_2)\right),$$
$$\int d\Sigma_j\{\ast\}=
\int du_j\tilde{V}(u_j)\int\limits_0^\infty 
d\alpha_je^{-{\alpha_j\over2}M^2}
\int D\xi_j\exp\left\{-\int\limits_0^{\alpha_j}
d\tau{\dot{\xi}_j^2(\tau)\over2}\right\}\{\ast\},$$
$$(j=1,2;~~~u_1=u,~~u_2=v),$$
$$ \xi_1(0)=v,~~~~~~\xi_1(\alpha_1)=x+u,~~~~~~~~~~
\xi_2(0)=-v,~~~~~~\xi_2(\alpha_2)=x-u.$$
where for simplicity we put $y=0$.

Our task is to get the asymptotic behaviour of the functions $G_P(x)$
for asymptotically large $x=\sqrt{x^2}\rightarrow\infty$. To this end,
let us introduce
the following variables:
$$ \alpha_j={x\over Ms_j},~~~~~~\tau_j={\beta_j\over Ms_j}.$$
$$ \xi_1(\beta)=n\beta+\eta_1(\beta),~~~~~~~
\xi_2(\beta)=n\beta+\eta_2(\beta),~~~~~n_{\mu}={x_{\mu}\over x}.$$
Then, one can obtain
\begin{eqnarray}
\label{pot3}
&& G_P(x)=\left({M\over8\pi^2x}\right)^2\int du\tilde{V}(u)
\int dv\tilde{V}(v)\\
&& \cdot\int\limits_{0}^{\infty}\int\limits_{0}^{\infty}ds_1 ds_2
\exp\left\{-{xM\over2}\left({1\over s_1}+s_1+
{1\over s_2}+s_2\right)\right\}
\cdot J_P(s_1,s_2;x),\nonumber
\end{eqnarray}
\begin{eqnarray}
\label{pot4}
&& J_P(s_1,s_2;x)\\
&& =\int\int D\eta_1 D\eta_2\exp\left\{-\int\limits_{0}^{x}d\beta
\left[{Ms_1\dot{\eta}_1^2(\beta)\over2}+
{Ms_2\dot{\eta}_2^2(\beta)\over2}\right]+W_x[\eta_1,\eta_2]\right\},
\nonumber\\
&&  \eta_1(0)=v,~~~~\eta_1(x)=u,~~~~~~~~~~\eta_2(0)=-v,
~~~~\eta_2(x)=-u,\nonumber
\end{eqnarray}
where
$$ W_x[\eta_1,\eta_2]=W_{11}+2W_{12}+W_{22},$$
$$ W_{ij}={g^2\over8M^2s_is_j}\int\limits_{0}^{x}\int\limits_{0}^{x}
d\beta_1 d\beta_2
D_m\left(n(\beta_1-\beta_2)+\eta_i(\beta_1)-\eta_j(\beta_2)\right),$$
The functional integral for $J_{x}(s_1,s_2;x)$ looks like the
Feynman path integral in the nonrelativistic statistic quantum mechanics
for the four-dimensional motion of particles $\eta_1(\beta)$ and
$\eta_2(\beta)$ with "masses" $Ms_1$ and $Ms_2$ where $\beta$ plays
the role of imaginary time or temperature. The interaction of
these particles is defined by the nonlocal functional
$W_{11}+2W_{12}+W_{22}$ which contains potential $W_{12}$ and
nonpotential $W_{11}+W_{22}$ interactions.

The asymptotic form of the function $J_P(s_1,s_2;x)$ looks like
\begin{eqnarray}
\label{pot5}
J_P(s_1,s_2;x)\sim\exp\{-x E_P(s_1,s_2)\},
\end{eqnarray}
where $E_P(s_1,s_2)$ is the energy of the lowest bound state.
The asymptotic behaviour of the functional $G_P(x)$ as $x\rightarrow\infty$
is determined by a saddle point of the integrals over $s_1$ and $s_2$
in the representation (\ref{pot3}). Substituting expression
(\ref{pot5}) into (\ref{pot3}), one can get
\begin{eqnarray}
\label{pot6}
\mu_P&=&-\lim_{x\to\infty}{1\over\vert x\vert}\ln G_P(x),\\
&=&\min_{(s_1, s_2)}\left[{M\over2}\left({1\over s_1}+s_1
+{1\over s_2}+s_2\right)+E_P(s_1,s_2)\right]\nonumber\\
&=&\min_s\left[M\left({1\over s}+s\right)+E_P(s,s)\right].\nonumber
\end{eqnarray}

The main problem is to compute the functional integral (\ref{pot4}).
There is not much hope, at least in the near future, that the
functional integral of the type (\ref{pot4}) will be calculetaed
exactly. We can only expect approximate methods to be worked out
to compute integrals of this kind with acceptable accuracy
especially for asymptotically large $x$.
We plan to calculate this functional integral applying the
Gaussian equivalent representation method (see \cite{Dineykhan}),
which was sucessfully used sucessfully for the polaron problem.

Besides one can see that this method is rough enough because it
does not really feel the explicit form of the vertex $\tilde{V}_Q(u)$
although it should extract a bound state with definite quantum
numbers $Q$.
It means, in fact, that in the general case, i.e., for any value of
the coupling constant $g$, the modern analytical methods, applied
to the functional integral (\ref{pot4}), allow one to calculate with
reasonable accuracy the energy of the lowest bound state only
(see \cite{Dineykhan}).

\subsection{The Green function $G_A(x-y)$}

The function $G_A(x)$ after integration over $\phi$ has the form
$$ G_A(x)=\left({M\over8\pi^2x}\right)^2
\int d\Sigma_1 d\Sigma_2\exp\{W_{11}+2W_{12}+W_{22}\},$$
$$ W_{jj}={g^2\over2}\int\limits_{0}^{\alpha_j}d\tau_1
\int\limits_{0}^{\alpha_j}d\tau_2
D_m\left(\xi_j(\tau_1)-\xi_j(\tau_2)\right),~~~~~(j=1,2),$$
$$ W_{12}={g^2\over2}\int\limits_{0}^{\alpha_1}d\tau_1
\int\limits_{0}^{\alpha_2}d\tau_2
D_m\left(x-y+\xi_1(\tau_1)-\xi_2(\tau_2)\right).$$
Here
$$ \xi_1(\tau)=x-u\left(1-{2\tau\over\alpha_1}\right)
+\int\limits_\tau^{\alpha_1}d\tau'\mu_1(\tau'),$$
$$ \xi_2(\tau)=y-v\left(1-{2\tau\over\alpha_2}\right)
+\int\limits_\tau^{\alpha_2}d\tau'\mu_2(\tau'),$$
$$ \xi_1(0)=-u,~~~~~\xi_1(\alpha_1)=u,~~~~~
\xi_2(0)=-v,~~~~~\xi_2(\alpha_2)=v.$$
In the following we put $y=0$ for simplicity.
To study the asymptotic behaviour of the functions $G_j(x)$
as $x=\sqrt{x^2}\rightarrow\infty$, we introduce the variables
$$ \alpha_j={x\over Ms_j},~~~~~~\tau_j={\beta_j\over Ms_j}.$$
One can obtain
\begin{eqnarray}
\label{pot7}
&& G_A(x)=\left({M\over8\pi^2x}\right)^2\int du\tilde{V}(u)
\int dv\tilde{V}(v)\\
&& \cdot\int\limits_{0}^{\infty}\int\limits_{0}^{\infty}ds_1 ds_2
\exp\left\{-{xM\over2}\left({1\over s_1}+s_1+{1\over s_2}+
s_2\right)\right\}
\cdot J_A(s_1,s_2;x),\nonumber
\end{eqnarray}
\begin{eqnarray}
&& J_A(s_1,s_2;x)\\
&& =\int\int D\eta_1 D\eta_2\exp\left\{-\int\limits_{0}^{x}d\beta
\left[{Ms_1\dot{\eta}_1^2(\beta)\over2}+
{Ms_2\dot{\eta}_2^2(\beta)\over2}\right]+W_x[\eta_1,\eta_2]\right\},
\nonumber
\end{eqnarray}
where
$$ W_x[\zeta_1,\zeta_2]=W_{11}+2W_{12}+W_{22},$$
$$ W_{jj}={g^2\over8M^2s_j^2}\int\limits_{0}^{x}\int\limits_{0}^{x}
d\beta_1 d\beta_2
D_m\left(\zeta_j(\beta_1)-\zeta_j(\beta_2)\right),~~~~(j=1,2),$$
$$ W_{12}={g^2\over8M^2s_1s_2}\int\limits_{0}^{x}\int\limits_{0}^{x}
d\beta_1 d\beta_2D_m\left(x+\zeta_1(\beta_1)-\zeta_2(\beta_2)\right),$$
$$ \zeta_1(0)=-u,~~~~~\zeta_1(x)=u,~~~~~\zeta_2(0)=-v,
~~~~~\zeta_2(x)=v.$$

One can see that in the limit $x\to\infty$
$$ W_{12}\to 0$$
and
$$ G_A(x)\to\left[{M\over8\pi^2x}\int d\Sigma_1e^{W_{11}}\right]^2,$$
so that no bound state arise in this case.
Thus, the annihilation channel does not contain any bound states. In
other words, intermediate pure boson states can not arange any bound
state.

\subsection{ The Nonrelativistic Limit }

In this section we obtain the nonrelativistic limit $c\rightarrow\infty$
for the loop function $G_P(x)$ in (\ref{pot3}-\ref{pot4}). To this end,
let us restore the parameter $c$ in our formulas:
$$ M\to Mc,~~~~~~g\to{g\over c},~~~~~~
x_\mu=(x_4,\vec{x})\to(ct,\vec{x}),$$
\begin{eqnarray}
&& D(x)\to cD(x)\nonumber\\
&&=c\int{d^4 k\over{(2\pi)^4}} \tilde{D}(k^2)e^{ikx}
=\int {d\vec{k}\over{(2\pi)^3}}
\int\limits_{-\infty}^{\infty}{dv\over{2\pi}}
\tilde{D}\left(\vec{k}^2+{v^2\over{c^2}}\right)e^{i(vt+\vec{k}\vec{x})}
\nonumber\\
&&=\int {d\vec{k}\over{(2\pi)^3}}
\int\limits_{-\infty}^{\infty}{dv\over{2\pi}}
\cdot{e^{i(vt+\vec{k}\vec{x})}\over\vec{k}^2+{v^2\over{c^2}}+\kappa^2}.
\nonumber
\end{eqnarray}
where $\kappa=mc$.

In the limit $c\to\infty$ we have
$$ x=\sqrt{x^2}=\sqrt{c^2t^2+\vec{x}^2}\rightarrow ct,~~~~~~
n_\mu=(1,{\vec x}/ct)\to (1,0),$$
Our task is to introduce the parameter $c$ in the explicit form into
$G_P(x)$ (\ref{pot7}) and then find the limit $c\rightarrow\infty$
in this expression for $G_P(x)$. Let us introduce in (\ref{pot7}) new
variables $\beta_j\to c\beta_j$. We obtain
\begin{eqnarray}
\label{non1}
G_P(t)&=&\left({M\over4\pi t}\right)^2
\int\limits_0^\infty\!\!\int\limits_0^\infty
ds_1ds_2\exp\left\{-t{Mc^2\over 2}\left({1\over s_1}+s_1
+{1\over s_2}+s_2\right)\right\}\nonumber\\
&\cdot&J_P(s_1,s_2;t),
\end{eqnarray}
\begin{eqnarray}
J_P(s_1,s_2;t)&=&\int\!\!\int D\eta_1D\eta_2
\exp\left\{-\int\limits_{0}^{t} d\beta\left[{Ms_1\dot{\eta}_1^2(\beta)\over2}
+{Ms_1\dot{\eta}_2^2(\beta)\over2}\right]\right\}\nonumber\\
&\cdot&\exp\left\{W_t[\eta_1,\eta_2]\right\},\nonumber
\end{eqnarray}
The functionals $W_{ij}$ become the form
\begin{eqnarray}
\label{non2}
&& W_{ij}={g^2\over8M^2s_is_j}\int\limits_{0}^{t}\int\limits_0^t
d\beta_1 d\beta_2 \int{d\vec{k}\over{(2\pi)^3}}
\int\limits_{-\infty}^\infty
{dv\over{2\pi}}\tilde{D}_m\left(\vec{k}^2+{v^2\over c^2}\right)\\
&& \cdot\exp\left\{iv\left((\beta_1-\beta_2)+{1\over c}
(\eta_{i4}(\beta_1) - \eta_{j4}(\beta_2))\right)
+i\vec{k}(\vec{\eta}_i(\beta_1)-\vec{\eta}_j(\beta_2))\right\}.\nonumber
\end{eqnarray}
In the case of the standard scalar propagator one can get
$$ \tilde{D}_m\left(\vec{k}^2+{v^2\over c^2}\right)=
{1\over\vec{k}^2+{v^2\over c^2}+\kappa^2}\to{1\over\vec{k}^2+\kappa^2}. $$
Here the parameter ${1\over\kappa}$ defines the radius of the Yukawa
potential; therefore, we will keep $\kappa=mc={\rm const}$ in the limit
$c\to\infty$.

Now we are able to go to the limit $c\rightarrow\infty$.
We can put all terms with ${1\over c}$ equal to zero. The
$\delta$-function  $\delta(\beta_1-\beta_2)$ arises because
the fourth components $\eta_{j4}(\beta)$ disappear in the
integrand. Thus, we get
\begin{eqnarray}
W_{ij}&\to&{g^2\over8M^2s_is_j}\int\limits_0^t\int\limits_0^t
d\beta_1 d\beta_2\delta(\beta_1-\beta_2)
\int{d\vec{k}\over{(2\pi)^3}}\tilde{D}_m\left(\vec{k}^2\right)
e^{i\vec{k}(\vec{\eta}_i(\beta_1)-\vec{\eta}_j(\beta_2))}\nonumber\\
&=&{1\over s_is_j}\int\limits_0^td\beta
U\left(\vec{\eta}_i(\beta)-\vec{\eta}_j(\beta)\right).\nonumber
\end{eqnarray}
where
$$ U(\vec{r})={g^2\over8M^2}\int{d\vec{k}\over{(2\pi)^3}}\tilde{D}_m
\left(\vec{k}^2\right)e^{i\vec{k}\vec{r}}
={\lambda\over8}\cdot{e^{-\kappa r}\over r}$$
is the nonrelativistic potential.

We shall consider only the term $W_{12}$, because the terms $W_{11}$ and
$W_{22}$ are equal to
$$ W_{jj}={t\over s_j^2}U(0)={g^2t\over8M^2s_j^2}
\int{d\vec{k}\over{(2\pi)^3}}\tilde{D}_m\left(\vec{k}^2\right) $$
and contribute to the mass renormalization of particles as well as
have purely relativistic origin.

In the limit $c\to\infty$ in the functional integral for
$J_P(s_1,s_2;t)$ we can integrate over the fourth components of
the 4-vectors $\eta_1$ and $\eta_2$
$$ \int D\eta_4\exp\left\{-\int\limits_{0}^{t}d\tau
{Ms\dot{\eta}_4^2(\tau)\over2}\right\}=
\exp\left\{-{Ms\over2}\cdot{(v_4-u_4)^2\over t}\right\}
\stackrel{t\to\infty}{\to}1$$
where $\eta_{j4}(0)=u_4,~~~~~~\eta_4(t)=v_4$.

Now we can integrate over the fourth components of
the 4-vectors $u$ and $v$:
\begin{eqnarray}
\label{non3}
&& \int du_4\tilde{V}(u)=\int du_4\tilde{V}(u_4,\vec{u})
=\psi(\vec{u}).
\end{eqnarray}

The next point is that the saddle points of the integrals over
$s_1$ and $s_2$, which determine the asymptotic behaviour of
the functional $G_P(x)$ as $x\rightarrow\infty$
in the limit $c\to\infty$, are $s_1=s_2=1$.
Thus, the functional integral for $J_P(t)=J_P(1,1;t)$ becomes
\begin{eqnarray}
\label{non4}
&& J_P(t)=e^{-t2Mc^2}\int d\vec{u}\psi(\vec{u})
\int d\vec{v}\psi(\vec{v})K(\vec{v},t;\vec{u},0),\\
&& K(\vec{v},t;\vec{u},0)=\int\!\!\!\int D\vec{\eta}_1D\vec{\eta}_2
\nonumber\\
&& \cdot\exp\left\{-\int\limits_0^t d\beta\left[
{M\dot{\vec{\eta}}_1^2(\beta)\over2}
+{M\dot{\vec{\eta}}_2^2(\beta)\over2}-
U\left(\vec{\eta}_1(\beta)-\vec{\eta}_2(\beta)\right)\right]\right\}.
\nonumber\\
&& \vec{\eta}_1(0)=\vec{u},~~~~\vec{\eta}_2(0)=-\vec{u},~~~~~
\vec{\eta}_1(t)=\vec{v},~~~~\vec{\eta}_2(t)=-\vec{v}.\nonumber
\end{eqnarray}
One can see that this representation for the function
$K(\vec{v},t;\vec{u},0)$ coincides with the Feynman path
integral in the quantum mechanics \cite{Feynman} for the situation
where there are two particles $\vec{\eta}_1$ and $\vec{\eta}_2$
with masses $M$ and the interaction between these particles
is described by the potential $U(\vec{\eta}_1-\vec{\eta}_2)$.

Let us introduce the standard variables
$$ \vec{\xi}_1(\beta)=\vec{R}(\beta)+{1\over2}\vec{r}(\beta),~~~~~~~
\vec{\xi}_2(\beta)=\vec{R}(\beta)-{1\over2}\vec{r}(\beta),$$
$$ \vec{R}(0)=\vec{R}(t)=0,~~~~~~~
\vec{r}(0)=2\vec{u},~~~~\vec{r}(t)=2\vec{v} $$
and integrate over $\vec{R}(\beta)$. We obtain
\begin{eqnarray}
\label{non5}
K(\vec{v},t;\vec{u},0)
&=&\int D\vec{r}
\exp\left\{-\int\limits_0^td\beta\left[{M_r\over2}\dot{\vec{r}}^2(\beta)
-V(\vec{r}(\beta))\right]\right\},\nonumber\\
&=&\sum\limits_N\psi_N(\vec{v})e^{-tE_N}\psi_N(\vec{u})
\end{eqnarray}
where the boundary conditions are $\vec{r}(0)=2\vec{u},
~~\vec{r}(t)=2\vec{v}$ and $M_r={M\over2}$.
Here $\psi_N(\vec{r})$ and $E_N$ are eigenfunctions
and eigenvalues for the quantun number $N$ connected with the space
$R^3$ of the Schr\"{o}dinger equation
\begin{eqnarray}
\label{non6}
\left[{p^2\over 2M_r}-U(\vec{r})\right]
\psi_N(\vec{r})=E_N\psi_N(\vec{r})
\end{eqnarray}
where $U(\vec{r})>0$, i.e., it is the attractive potential.

As a result, the function $J_P(t)$ for $t\to\infty$ behaves like
\begin{eqnarray}
&& J_P(t)=\sum\limits_Ne^{-tE_N}
\left[\int d\vec{u}\psi(\vec{u})\psi_N(\vec{u})\right]^2.\nonumber
\end{eqnarray}
If we choose
$$\psi(\vec{u})=\psi_{N_0}(\vec{u})$$
then
$$\int d\vec{u}\psi_N(\vec{u})\psi_{N_0}(\vec{u})
=\delta_{NN_0}$$
and, finally, for large $t$ we have
\begin{eqnarray}
&& J_P(t) \rightarrow e^{-tE_{N_0}}.\nonumber
\end{eqnarray}
where $E_{N_0}$ is the energy of the bound state of two
nonrelativistic particles in the quantum state $N_0$ arising
due to the potential $U(\vec{r})$.
The mass of the bound state in the nonrelativistic approach is
\begin{eqnarray}
\label{non7}
\mu_{N_0}&=&{1\over c^2}\left[Mc^2+E_{N_0}+
O\left({1\over c}\right)\right]
=2M+{E_{N_0}\over c^2}+O\left({1\over c^3}\right).
\end{eqnarray}
Thus, in the nonrelativitic limit the mass of
the bound state of two scalar particles is the sum of their masses
plus the bound state energy which is defined by the nonrelativistic
potential interaction.

\subsection{ The Nonrelativistic Limit. Dimensional analysis.}

Here we show that the nonrelativistic limit is completely equivalent
to the weak coupling regime when $\lambda\to0$.

The dimension of our variables and parameters is
$$[M]=[g]=\left[{1\over x}\right]=\left[{1\over\beta_j}\right]=
[\eta_j].$$
Let us consider three functions in formula (\ref{pot3}) for $G_P(x)$
\begin{eqnarray}
A&=&{xM\over2}\left({1\over s_1}+s_1+{1\over s_2}+s_2\right),\nonumber\\
K&=&\int\limits_{0}^{x}d\beta\left[{Ms_1\dot{\eta}_1^2(\beta)\over2}+
{Ms_2\dot{\eta}_2^2(\beta)\over2}\right],\nonumber\\
W_{ij}&=&{g^2\over8M^2s_is_j}\int\limits_{0}^{x}\int\limits_{0}^{x}
d\beta_1 d\beta_2 D\left(n(\beta_1-\beta_2)+
\eta_i(\beta_1)-\eta_j(\beta_2)\right)\nonumber
\end{eqnarray}
where $n=(1,\vec{0})$ and
$$ D(x)=\int{dk\over(2\pi)^4}\tilde{D}(k^2)e^{ikx}=
\int{d\vec{k}\over(2\pi)^3}\int{dk_4\over2\pi}
\tilde{D}(\vec{k}^2+k_4^2)e^{i(k_4x_4+\vec{k}\vec{x})}.$$
Let us introduce the following dimensionless variables:
$$ \beta_j={M\over g^2}\tau_j,~~~~~\eta_j={1\over g}\zeta_j,
~~~~~~x={1\over M}\left({M\over g}\right)^2\cdot X,$$
$$ \vec{k}=g\vec{q},~~~~~k_4={g^2\over M}v.$$
One obtains
\begin{eqnarray}
&& A=X\left({M\over g}\right)^2\cdot
\left({1\over s_1}+s_1+{1\over s_2}+s_2\right),\nonumber\\
&& K=\int\limits_{0}^{X}d\tau\left[{s_1\dot{\zeta}_1^2(\tau)\over2}+
{s_2\dot{\zeta}_2^2(\tau)\over2}\right],\nonumber\\
&& W_{ij}={g\over M}{1\over8s_is_j}
\int\limits_{0}^{X}\int\limits_{0}^{X}
d\tau_1 d\tau_2 \int{d\vec{q}\over(2\pi)^3}\int{dv\over2\pi}
g^2\tilde{D}\left(g^2\left[\vec{q}^2+{g^2\over M^2}v^2\right]\right)
\nonumber\\
&& \cdot\exp\left\{iv\left[(\tau_1-\tau_2)+{g\over M}(\zeta_{i4}
(\tau_1)-\zeta_{j4}(\tau_2))\right]+
i\vec{q}(\vec{\zeta}_i(\tau_1)-\vec{\zeta}_j(\tau_2))\right\}.\nonumber
\end{eqnarray}
In particular for
$$\tilde{D}(k^2)={1\over k^2+m^2}$$
we get
$$ g^2\tilde{D}\left(g^2\left[\vec{q}^2+{g^2\over M^2}v^2\right]\right)=
{1\over\vec{q}^2+{g^2\over M^2}v^2+{m^2\over g^2}}.$$

In the weak coupling regime $\lambda\ll 1$ we have
\begin{eqnarray}
&& \int{dv\over2\pi}
g^2\tilde{D}\left(g^2\left[\vec{q}^2+{g^2\over M^2}v^2\right]\right)
e^{iv\left[(\tau_1-\tau_2)+{g\over M}(\zeta_{i4}
(\tau_1)-\zeta_{j4}(\tau_2))\right]}\nonumber\\
&& \to\int{dv\over2\pi}
g^2\tilde{D}\left(g^2\vec{q}^2\right)e^{iv(\tau_1-\tau_2)}=
g^2\tilde{D}\left(g^2\vec{q}^2\right)\delta(\tau_1-\tau_2)\nonumber
\end{eqnarray}
and
\begin{eqnarray}
W_{ij}&\to&{g\over M}{1\over8s_is_j}
\int\limits_{0}^{X}d\tau \int{d\vec{q}\over(2\pi)^3}
g^2\tilde{D}\left(g^2\vec{q}^2\right)
e^{i\vec{q}(\vec{\zeta}_i(\tau)-\vec{\zeta}_j(\tau))}\nonumber\\
&=&{1\over s_is_j}\int\limits_{0}^{X}d\tau
U(\vec{\zeta}_i(\tau)-\vec{\zeta}_j(\tau)).
\end{eqnarray}
where
\begin{eqnarray}
U(\zeta)&=&{g\over 8M}\int{d\vec{q}\over(2\pi)^3}
g^2\tilde{D}\left(g^2\vec{q}^2\right)e^{i\vec{q}\vec{\zeta}},\nonumber\\
&=&{g\over 8M}\int{d\vec{q}\over(2\pi)^3}
{e^{i\vec{q}\vec{\zeta}}\over\vec{q}^2+{m^2\over g^2}}=
{g\over M}{1\over32\pi}
{e^{-{m\over g}\zeta}\over\zeta}\nonumber\\
&=& M{\lambda\over 8}{e^{-m\eta}\over\eta},
\nonumber
\end{eqnarray}
where $\zeta=\vert\vec{\zeta}\vert$ and $\eta=\vert\vec{\eta}\vert$.

Thus, nonrelativistic limit is realized in the weak coupling regime.

\subsection{Relativistic incompleteness
of quantum mechanics of two particles.}

Here we would like to pay attention to the Schr\"{o}dinger equation
describing two nonrelativistic particles
\begin{eqnarray}
&& H={p_1^2\over2m_1}+{p_2^2\over2m_2}-U(\vec{r}_1-\vec{r}_2),
\nonumber
\end{eqnarray}
where the potential is attractive.

Let us pass to the inertial system in the standard way
$$ \vec{R}={m_1\vec{r}_1+m_2\vec{r}_2\over m_1+m_2},~~~~~~~
\vec{r}=\vec{r}_1-\vec{r}_2. $$
The Hamiltonian takes the form
\begin{eqnarray}
&& H={p^2\over2M}+{p_r^2\over2\mu}-U(r)\nonumber
\end{eqnarray}
$$ M=m_1+m_2,~~~~~~~\mu={m_1m_2\over m_1+m_2} $$
The solution of the Schr\"{o}dinger equation
\begin{eqnarray}
&& H\Psi=E\Psi\nonumber
\end{eqnarray}
can be written as
$$ \Psi(\vec{R},\vec{r})=e^{i\vec{p}\vec{R}}\psi(r)$$
where $\vec{p}$ is the momentum of the total system and
$\psi(r)$ is the eigenfunction of the equation
\begin{eqnarray}
&& \left[{p_r^2\over2\mu}-U(r)\right]\psi=-\varepsilon\psi
\nonumber
\end{eqnarray}
and $-\varepsilon~(\varepsilon>0)$ is an eigenvalue of a bound state.
Then, the eigenvalue or the energy of the state $\Psi(\vec{R},\vec{r})$
in the case $\vec{p}\neq 0$ is
\begin{eqnarray}
&& E={\vec{p}^2\over2M}-\varepsilon.\nonumber
\end{eqnarray}
>From the physical point of view this energy has no reasonable sense.
Indeed, we should get
\begin{eqnarray}
&& E={p^2\over2M}-\varepsilon\to{p^2\over2M_{phys}},\nonumber\\
&& M_{phys}=m_1+m_2-\Delta,~~~~~~~\Delta={\varepsilon\over c^2}\nonumber
\end{eqnarray}
i.e., the interaction between two particles should give the mass excess.

On the other hand, the last formula can be obtained from the pure
energy in the nonrelativistic limit
\begin{eqnarray}
E&=&\sqrt{M_{phys}^2c^4+p^2c^2}=M_{phys}c^2+{p^2\over2M_{phys}}+O(p^2)
\nonumber\\
&=&(m_1+m_2-\Delta)c^2+{p^2\over(m_1+m_2-\Delta)}+O(p^2)\nonumber\\
&=&(m_1+m_2)c^2+{p^2\over2(m_1+m_2)}-\Delta c^2+
O\left({\Delta\over M}\right)+O(p^2),\nonumber
\end{eqnarray}
and the mass of the bound state
\begin{eqnarray}
&& M_{phys}=Mc^2-{\varepsilon\over c^2}+O\left({1\over c^2}\right).
\nonumber
\end{eqnarray}
Thus, we can consider the nonrelativistic Schr\"{o}dinger equation
describing two nonrelativistic particles as a fragment of an unknown
relativistic equation describing the relativistic interaction of
two particles.

\section{Bosonization of fermion currents.}

The starting point of the Bosonization of fermion currents
is the representation (\ref{for9}).

\subsection{Fierz transformation.}

Let us consider the four fermion term (\ref{for10}) and perform the "Fierz
transformation":
\begin{eqnarray}
\label{bos1}
&& (\Psi^+\Psi D_m\Psi^+\Psi)=\int\int dy_1dy_2\Psi^+(y_1)\Psi(y_1)
D_m(y_1-y_2)\Psi^+(y_2)\Psi(y_2)\nonumber\\
&&=-\int\int dy_1dy_2\Psi^+(y_1)\Psi(y_2)
D_m(y_1-y_2)\Psi^+(y_2)\Psi(y_1).
\end{eqnarray}
Let us introduce the new variables
$$ y_1=x+{u\over 2},~~~~~~~y_2=x-{u\over 2}$$
Then the four fermion term looks like
\begin{eqnarray}
\label{bos2}
(\Psi^+\Psi D_m\Psi^+\Psi)=-\int dx\int du D_m(u)J(x,u)J^+(x,u)
\end{eqnarray}
where
\begin{eqnarray}
&& J(x,u)=
\left(\Psi^+\left(x+{u\over2}\right)\Psi\left(x-{u\over2}\right)\right)=
\Psi^+(x)e^{{u\over2}\stackrel{\leftrightarrow}{\partial}_x}\Psi(x),\\
&& J^+(x,u)=J(x,-u).\nonumber
\end{eqnarray}

\subsection{Orthonormal system.}

The main point is that in the representation (\ref{bos2}) the boson
Green function $D_m(u)$ can be considered as a weight function
inducing uniquely the system of orthogonal polynomials in the space $R^4$.
Thus, the full orthonormal system of real functions
$$\{f^{(nl)}_{\{\mu_1...\mu_l\}}(u)\}=\{f^{(nl)}_{\{\mu\}}(u)\}
=\{f_Q(u)\},~~~~~~~Q={nl\{\mu\}},$$
which are symmetric for all permutations
$\mu_i\rightleftharpoons\mu_j$ and
\begin{eqnarray}
&& f^{(nl)}_{\{\nu\nu\mu_3...\mu_l\}}(u)\equiv0,\nonumber
\end{eqnarray}
can be chosen in the form
\begin{eqnarray}
\label{bos3}
&& f_Q(u)=\sqrt{\rho(u)}P_Q(u),\\
&& \rho(u)=D_m(u)=\int{dk\over(2\pi)^4}{e^{iku}\over m^2+k^2}=
{m^2\over(2\pi)^2}\cdot{1\over mu}K_1(mu),~~~~~u=\sqrt{u^2}\nonumber
\end{eqnarray}
where $P_Q(u)$ are real polynomials, satisfying
$$ P_Q(-u)=(-1)^lP_Q(u).$$

The orthonormality conditions look like
\begin{eqnarray}
(f_Qf_{Q'})&=&\int d^4uf_Q(u)f_{Q'}(u)
=\int d^4u\rho(u)P_Q(u)P_{Q'}(u)\nonumber\\
&=&\delta_{QQ'}=\delta_{nn'}\delta_{ll'}\delta_{\{\mu\}\{\mu'\}},
\nonumber
\end{eqnarray}
$$ \rho(u)\sum_QP_Q(u)P_Q(u')=\delta(u-u')$$
The symbol $\delta_{\{\mu\}\{\mu'\}}$ is defined as
$$ \sum_{\{\mu'\}}\delta_{\{\mu\}\{\mu'\}}f^{(nl)}_{\{\mu'\}}(u)=
f^{(nl)}_{\{\mu\}}(u).$$
The orthonormal condition can be rewritten in the form
\begin{eqnarray}
\label{bos4}
&& (f_Qf_{Q'})=\left.P_Q\left(i{\partial\over\partial k}\right)
P_{Q'}\left(i{\partial\over\partial k}\right)
{1\over m^2+k^2}\right\vert_{k=0}=\delta_{QQ'}.
\end{eqnarray}
For the lowest states one can get
\begin{eqnarray}
&& P_{00}=m,~~~~~~~~~P_{10}(u)={m\over\sqrt{2}}\left(1-{m^2u^2\over8}\right),
\nonumber\\
&& P_{01\{\mu\}}(u)={m^2\over\sqrt{2}}u_\mu,~~~~~~~~~~~~~
P_{11\{\mu\}}(u)={m^2\over\sqrt{2}}\left(1-{m^2u^2\over24}\right)u_\mu,
\nonumber\\
&& P_{02\{\mu\nu\}}(u)={m^3\over4}\left(u_\mu u_\nu-
{1\over4}\delta_{\mu\nu}u^2\right),~~~~~{\rm and~~so~~on}.\nonumber
\end{eqnarray}
The normalization conditions for these states look like
\begin{eqnarray}
&& (f_{00}f_{00})=(f_{10}f_{10})=1,\nonumber\\
&& (f_{01\{\mu\}}f_{01\{\nu\}})=
(f_{11\{\mu\}}f_{11\{\nu\}})=\delta_{\mu\nu},\nonumber\\
&& (f_{02\{\mu\nu\}}f_{02\{\alpha\beta\}})=
\delta_{\{\mu\nu\}\{\alpha\beta\}}={1\over2}\left[
\delta_{\mu\alpha}\delta_{\nu\beta}+\delta_{\mu\beta}\delta_{\nu\alpha}
-{1\over2}\delta_{\mu\nu}\delta_{\alpha\beta}\right].\nonumber
\end{eqnarray}

Let us introduce the vertex
\begin{eqnarray}
\label{bos5}
V_Q(p)&=&(-i)^l\int duf_Q(u)\sqrt{\rho(u)}e^{iup}
=(-i)^l\int duP_Q(u)\rho(u)e^{iup}\nonumber\\
&=&(-i)^lP_Q\left(-i{\partial\over\partial p}\right){1\over m^2+p^2},\\
&& V_Q^+(p)=V_Q(p),~~~~~~~~V_Q^+(-p)=(-1)^lV_Q(p).\nonumber
\end{eqnarray}
In particular
\begin{eqnarray}
&& V_{00}(p)={m\over m^2+p^2},~~~~~~~~
V_{01\{\mu\}}(p)={m^2\over\sqrt{2}}\cdot{2p_\mu\over(m^2+p^2)^2},\nonumber\\
&& V_{10}(p)=
{m\over\sqrt{2}}\cdot{p^2(2m^2+p^2)\over(m^2+p^2)^3}.\nonumber
\end{eqnarray}
Then, the representation is valid
\begin{eqnarray}
&& \sqrt{\rho(u)}e^{i{u\over2}p}=\sum_Qi^lf_Q(u)
V_Q\left({p\over2}\right).\nonumber
\end{eqnarray}

\subsection{Bilocal and nonlocal currents.}

The bilocal current $J(x,u)$ can be represented as
\begin{eqnarray}
\label{bos6}
\sqrt{D_m(u)}J(x,u)&=&\sqrt{\rho(u)}
(\Psi^+(x)e^{{u\over2}\stackrel{\leftrightarrow}{\partial}_x}\Psi(x))
=\sum_Qi^lf_Q(u)\cdot{\rm I}_Q(x),\nonumber\\
{\rm I}_Q(x)&=&\left(\Psi^+(x)V_Q\left({\stackrel{\leftrightarrow}{p}_x
\over2}\right)\Psi(x)\right)=\left(\Psi^+V_Q\Psi\right),\nonumber\\
&& {\rm I}_Q^+(x)={\rm I}_Q(x).\nonumber
\end{eqnarray}
Thus, the bilocal currents $J(x,u)$ are represented in the form of a sum
of the nonlocal hermitian currents ${\rm I}_Q(x)$.
As a result, the representation is valid
\begin{eqnarray}
\label{bos7}
&& (\Psi^+\Psi D_m\Psi^+\Psi)=-\sum_Q\int dx\left[{\rm I}_Q(x)\right]^2.
\end{eqnarray}

The next step is to use the Gaussian representation
\begin{eqnarray}
&& \exp\left\{{g^2\over2}(\Psi^+\Psi D_m\Psi^+\Psi)\right\}
=\exp\left\{-{g^2\over2}\sum_Q\int dx
{\rm I}_Q^2(x)\right\}\nonumber\\
&& =\int D\Phi\exp\left\{-{1\over2}(\Phi\Phi)+ig(\Phi{\rm I})\right\}.
\nonumber
\end{eqnarray}
where
\begin{eqnarray}
&& D\Phi=\prod_QD\Phi_Q,~~~~~~~~(\Phi\Phi)=\sum_Q\int dx\Phi_Q^2(x),
\nonumber
\end{eqnarray}
\begin{eqnarray}
(\Phi{\rm I})&=&\sum_Q\int dx\Phi_Q(x){\rm I}_Q(x)
=\sum_Q(\Psi^+[\Phi_QV_Q]\Psi)=(\Psi^+[\Phi V]\Psi).\nonumber
\end{eqnarray}
The hermitian field variables
\begin{eqnarray}
&& \Phi_Q=\Phi^{(nl)}_{\{\mu_1...\mu_l\}}(u)\nonumber
\end{eqnarray}
are symmetric for all permutations
$\mu_i\rightleftharpoons\mu_j$ and
\begin{eqnarray}
&& \Phi^{(nl)}_{\{\nu\nu\mu_3...\mu_l\}}(u)\equiv0,\nonumber
\end{eqnarray}

Let us suppose that $V=gV_{Q_0}$. For $x\neq y$ we have
\begin{eqnarray}
G_{Q_0}(x-y)&=&\int D\Psi D\Psi^+
~(\Psi^+V_{Q_0}\Psi)_x(\Psi^+V_{Q_0}\Psi)_y~
\exp\left\{(\Psi^+D^{-1}_M\Psi)\right\}\cdot\nonumber\\
&\cdot&\int D\Phi\exp\left\{-{1\over2}(\Phi\Phi)
+ig(\Psi{\rm I})\right\}\nonumber\\
&=&-\int D\Phi\exp\left\{-{1\over2}(\Phi\Phi)\right\}
{\delta^2\over\delta\Phi_{Q_0}(x)\delta\Phi_{Q_0}(y)}
\cdot\nonumber\\
&\cdot&\int D\Psi D\Psi^+\exp\left\{(\Psi^+D^{-1}_M\Psi)
+ig\int dx(\Psi^+[\Phi V]\Psi)\right\}\nonumber\\
&=&-\int D\Phi\exp\left\{-{1\over2}(\Phi\Phi)\right\}\nonumber\\
&\cdot&{\delta^2\over\delta\Phi_{Q_0}(x)\delta
\Phi_{Q_0}(y)}\exp\left\{{\rm tr}
\ln\left[1+ig(\Phi V)D_M\right]\right\}.
\nonumber
\end{eqnarray}
Integrating by parts we finally get
\begin{eqnarray}
\label{bos8}
&& G_{Q_0}(x-y)=-\int D\Phi~\Phi_{Q_0}(x)\Phi_{Q_0}(y)~
e^{L_{eff}[\Phi]},
\end{eqnarray}
where
\begin{eqnarray}
\label{bos9}
&& L_{eff}[\Phi]=-{1\over2}(\Phi\Phi)+
{\rm tr}\ln\left[1+ig(\Phi V)D_M\right].
\end{eqnarray}
First of all we would like to stress that the representation (\ref{bos8})
is completely equivalent to the initial representation (\ref{for9}).
The Green function $G_{Q_0}(x-y)$ can be considered as the Green
function of the field $\Phi_{Q_0}(x)$ and these fields $\{\Phi_{Q_0}\}$
are described by the effective Lagrangian $L_{eff}[\Phi]$. In addition,
we want to pay attention to the sign "minus" in front of the expression
(\ref{bos8}) and the nonhermitian interaction Lagrangian in (\ref{bos8})
and (\ref{bos9}).

\subsection{One-loop approach.}

The next problem is to give the standard particle interpretation to the
effective Lagrangian (\ref{bos9}). For this aim, let us extract
the quadratic term
\begin{eqnarray}
&& {\rm tr}\ln\left[1+ig(\Phi V)D_M\right]=
{g^2\over2}(\Phi\Pi\Phi)+{\rm I}_{int}[\Phi],\nonumber
\end{eqnarray}
\begin{eqnarray}
g^2(\Phi\Pi\Phi)&=&\sum_{QQ'}(\Phi_Qg^2[V_QD_M V_{Q'}D_M]\Phi_{Q'})
\nonumber\\
&=&\int dx\int dy\sum_{QQ'}
\Phi_Q(x)g^2\Pi_{QQ'}(x-y)
\Phi_{Q'}(y)\nonumber\\
&=&\int dp\sum_{QQ'}\tilde{\Phi}_Q(p)
g^2\tilde{\Pi}_{QQ'}(p)\tilde{\Phi}_{Q'}(p)\nonumber
\end{eqnarray}
where the polarization operator is
\begin{eqnarray}
\label{bos10}
&& g^2\Pi_{QQ'}(x-y)\\
&& =g^2\left[V_Q
\left({\stackrel{\leftarrow}{p}_x-\stackrel{\rightarrow}{p}_x\over2}\right)
D_M(x-y)V_{Q'}
\left({\stackrel{\leftarrow}{p}_y-\stackrel{\rightarrow}{p}_y\over2}\right)
D_M(y-x)\right],\nonumber\\
&& =g^2\int{dk\over(2\pi)^4}\tilde{D}_M(k)
\int{dk'\over(2\pi)^4}\tilde{D}_M(k')e^{i(x-y)(k-k')}\nonumber\\
&&\cdot\left[V_Q\left({k+k'\over2}\right)
V_{Q'}\left({k+k'\over2}\right)\right],\nonumber
\end{eqnarray}
and in the momentum space
\begin{eqnarray}
\label{bos11}
&& g^2\tilde{\Pi}_{QQ'}(p)=
\int dxe^{ipx}g^2\Pi_{QQ'}(x)
=g^2\tilde{\Pi}_{QQ'}(p).\nonumber\\
&& g^2\tilde{\Pi}_{QQ'}(p)
=g^2\int{dk\over(2\pi)^4}{V_Q(k)V_{Q'}(k)
\over\left(M^2+\left(k+{p\over2}\right)^2\right)
\left(M^2+\left(k-{p\over2}\right)^2\right)}\nonumber
\end{eqnarray}
The index structure of the polarization operator looks like
\begin{eqnarray}
&& \tilde{\Pi}_{QQ}(p)
=\tilde{\Pi}^{(nl)}_{\{\mu_1\}\{\mu_2\}}(p)
=\tilde{\Pi}_{(nl)}(p^2)\delta_{\{\mu_1\}\{\mu_2\}}+
\sum_jg^2\tilde{\Pi}^{(nl)}_j(p^2)t^j_{\{\mu_1\}\{\mu_2\}}(p)\nonumber
\end{eqnarray}
where the tensors $t^j_{\{\mu_1\}\{\mu_2\}}(p)$ contain the vectors
$p_\mu$.

The Green function $G_{Q_0}$ takes the form
\begin{eqnarray}
\label{bos12}
&& G_{Q_0}(x-y)\\
&&=-\int D\Phi~\Phi_{Q_0}(x)\Phi_{Q_0}(y)~
\exp\left\{-{1\over2}(\Phi[1-g^2\Pi]\Phi)+{\rm I}_{int}[\Phi]\right\},
\nonumber\\
&& (\Phi[I-g^2\Pi]\Phi)=\int dp\sum_{QQ'}
\tilde{\Phi}_Q(p)\left[\delta_{QQ'}-g^2\tilde{\Pi}_{QQ'}(p)\right]
\tilde{\Phi}_{Q'}(p),\nonumber
\end{eqnarray}
where
\begin{eqnarray}
{\rm I}_{int}[\Phi]={\rm tr}\ln\left
[1+ig(\Phi V)D_M\right]-{g^2\over2}(\Phi\Pi\Phi).\nonumber
\end{eqnarray}

The diagonal part of the quadratic form of (\ref{bos12}) gives the equation
of motion on the field $\Phi^{(nl)}_{a\{\nu\mu_2...\mu_l\}}(x)$
\begin{eqnarray}
&& \left[\delta_{QQ'}-g^2\tilde{\Pi}_{QQ'}
\left({\partial\over i\partial x}\right)\right]\Phi_{Q'}(x)=0,\nonumber
\end{eqnarray}
or
\begin{eqnarray}
\label{bos13}
&& \left[\delta_{QQ'}-g^2\tilde{\Pi}_{QQ'}(p)\right]
\tilde{\Phi}_{Q'}(p)=0.
\end{eqnarray}
The requirement that this equation on the mass shell should be
the Klein-Gordon equation gives the constrain
\begin{eqnarray}
&& {\partial\over\partial x_{\nu}}\Phi^{(nl)}_{\nu\mu_2...\mu_l}(x)=0
~~~~~~{\rm or}~~~~~~
p_\nu\tilde{\Phi}^{(nl)}_{\{\nu\mu_2...\mu_l\}}(p)=0\nonumber
\end{eqnarray}
on the mass shell. Then, the function
$\tilde{\Phi}^{(nl)}_{\{\mu\}}(p)$ satisfies the equation
\begin{eqnarray}
\label{bos14}
&& \left[1-\tilde{\Pi}^{(nl)}(p^2)\right]
\tilde{\Phi}^{(nl)}_{\{\mu_1...\mu_l\}}(p)=0.
\end{eqnarray}
The mass of the state with quantum numbers $Q=(nl)$ is
defined by the equation
\begin{eqnarray}
\label{bos15}
&& 1-g^2\tilde{\Pi}^{(nl)}(-\mu_{(nl)}^2)=0.
\end{eqnarray}
Let us represent
\begin{eqnarray}
\label{bos16}
1-g^2\tilde{\Pi}^{(nl)}(p^2)&=&[1-g^2\tilde{\Pi}^{(nl)}(-\mu_{(nl)}^2)]\\
&-&g^2\tilde{\Pi}_{(nl)}'(-\mu_{(nl)}^2)(p^2+\mu_{(nl)}^2)
-g^2\tilde{\Pi}^{(nl)}_{reg}(p^2,\mu_{(nl)}^2)\nonumber\\
&=&-Z_{(nl)}(p^2+\mu_{(nl)}^2)
-g^2\tilde{\Pi}^{(nl)}_{reg}(p^2,\mu_{(nl)}^2)\nonumber
\end{eqnarray}
where the constant
\begin{eqnarray}
&& Z_{(nl)}=g^2\tilde{\Pi}_{(nl)}'(-\mu_{(nl)}^2)\nonumber
\end{eqnarray}
is positive.

The crucial point is that in the representation (\ref{bos8}) the sign of
the measure of the functional integral is defined by
$$ e^{-{1\over2}(\Phi\Phi)} $$
but in the representation (\ref{bos12}) by
$$ e^{-{1\over2}(\Phi[1-g^2\Pi]\Phi)} $$
and according to (\ref{bos16}) by
$$ e^{{1\over2}(\Phi Z_{(nl)}[p^2+\mu_{(nl)}^2]\Phi)}. $$
We can see that the sign of the quadratic form becomes positive.
It requires to do the rotation
$$\Phi\to-i\phi $$
to have the decreasing measure. Then, let us introduce new field
variables
\begin{eqnarray}
&& \Phi_Q(x)={-i\over\sqrt{Z_{(nl)}}}\varphi_Q(x).\nonumber
\end{eqnarray}
Finally we get
\begin{eqnarray}
\label{bos17}
&& G_{Q_0}(x-y)={1\over Z_{Q_0}}{\cal G}_{Q_0}(x-y),\\
&&{\cal G}_{Q_0}(x-y)=\int D\varphi~\varphi_{Q_0}(x)\varphi_{Q_0}(y)~
e^{L_{eff}[\phi]}=D_{\mu_{(nl)}}(x-y)+O(h^2).\nonumber
\end{eqnarray}
The normalization constant in (\ref{bos17}) should be implied
$$ \varphi\to{D\varphi\over C},~~~~~~~C=\int D\varphi~e^{L_{eff}[\phi]}.$$
The effective Lagrangian looks like
\begin{eqnarray}
\label{bos18}
&& L_{eff}[\phi]={1\over2}(\varphi\left[\Box-\mu^2\right]\varphi)
+{\cal I}_{int}[\varphi]
\end{eqnarray}
Here
\begin{eqnarray}
(\varphi\left[-\Box+\mu^2\right]\varphi)&=&
\int dx\sum_{\cal Q}\varphi_Q(x)
\left[-\Box+\mu_{(nl)}^2\right]\varphi_Q(x)\nonumber\\
&=&\int dp\sum_{\cal Q}\tilde{\varphi}_Q^+(p)
\left[p^2+\mu_{(nl)}^2\right]\tilde{\varphi}_Q(p),\nonumber
\end{eqnarray}
\begin{eqnarray}
\label{bos19}
&& {\cal I}_{int}[\varphi]={\rm tr}\ln\left
[1+h\varphi VD_M\right]+(\varphi\Pi\varphi),\\
&& (\varphi\Pi\varphi)=\sum_{Q\neq Q'}
h_Qh_{Q'}(\varphi_Q\Pi_{QQ'}\varphi_{Q'})
+\sum_Qh_Q^2(\varphi_Q\Pi_Q^{reg}\varphi_Q),\nonumber\\
&& h\varphi V=\sum_Qh_Q\varphi_QV_Q,\nonumber\\
&& {\cal I}^+_{int}[\varphi]={\cal I}_{int}[\varphi].\nonumber
\end{eqnarray}
The effective coupling constants are defined as
\begin{eqnarray}
\label{boss19}
&& h_Q=h_{(nl)}={g\over\sqrt{Z_{(nl)}}}=
{1\over\sqrt{\tilde{\Pi}'_{(nl)}(-\mu_{(nl)}^2)}}.
\end{eqnarray}

Now we want to repeat and to stress that eq. (\ref{bos15}) and
the positivity of $Z_{(nl)}$ require the rotation $\Phi\to-i\phi$,
so that the sign $(-i)^2=-1$ arises in front of the Green function
${\cal G}_{Q_0}(x-y)$ and compensates the sign $(-1)$ in (\ref{bos8}).
Besides, the interaction Lagrangian becomes hermitian. Thus, we get the
representation (\ref{bos17}) which has the standard physical sense with
the positive metrics and the hermitian effective Lagrangian (\ref{bos18}).

As a result, the final representation (\ref{bos17}) can be interpreted
as a generating functional of the quantum field system of the
bosonic fields $\{\phi_Q\}$ which is described by the effective
Lagrangian (\ref{bos18}).

\subsection{The functions $\tilde{\Pi}_{(nl)}(p^2)$.}

Let us evaluate the values of $\lambda$ for some typical
parameters ${m\over M}$ and possible masses ${\mu\over2M}$ of the lowest
bound states with quantum number $(00),~(10),~(01)$. We have
\begin{eqnarray}
\label{sp1}
&& g^2\tilde{\Pi}_{(n0)}(p^2)\\
&&=g^2\int{dk\over(2\pi)^4}\cdot
{[V_{n0}(k^2)]^2\over\left(M^2+\left(k+{p\over2}\right)^2\right)
\left(M^2+\left(k-{p\over2}\right)^2\right)}\nonumber\\
&&=\lambda\cdot{\xi\over8\pi b}\cdot\int\limits_0^1dt
w_{(n)}(\xi,t)\cdot\left[{\sqrt{(1-bt)^2+4bt(1-t)}\over1-bt}-1\right]
\nonumber\\
&& w_{(0)}(\xi,t)={1\over(1-(1-\xi)t)^2},\nonumber\\
&& w_{(1)}(\xi,t)={1\over2}\left[{(1-t)(1-(1-2\xi)t)
\over(1-(1-\xi)t)^3}\right]^2,
\nonumber
\end{eqnarray}
where
$$ \xi=\left({m\over M}\right)^2,
~~~~~~~b=-{p^2\over4M^2}={\mu^2\over4M^2}.$$
For the state $(01)$ in the integrand in (\ref{bos11}) we have
$$ k_\mu k_\nu\to{\delta_{\mu\nu}\over3}
\left[k^2-{(kp)^2\over p^2}\right]+
{p_\mu p_\nu\over p^2}{1\over3}\left[-k^2+4{(kp)^2\over p^2}\right]$$
and for the diagonal part of the polarization operator
$\tilde{\Pi}_{(01)}(p^2)$ one can get
\begin{eqnarray}
\label{sp2}
&& g^2\tilde{\Pi}_{(01)}(p^2)\\
&&={g^2\over3}\int{dk\over(2\pi)^4}\cdot{2m^2\over(m^2+k^2)^4}\cdot
{k^2-{(kp)^2\over p^2}\over\left(M^2+\left(k+{p\over2}\right)^2\right)
\left(M^2+\left(k-{p\over2}\right)^2\right)}\nonumber\\
&&=\lambda\cdot{\xi^2\over48\pi b^2}\cdot\int\limits_0^1
{dt\over(1-(1-\xi)t)^4}\nonumber\\
&& \cdot
\left\{{[(1-bt)^2+4bt(1-t)]^{3/2}\over1-bt}-(1-bt)^2-6bt(1-t)\right\}
\nonumber
\end{eqnarray}
It is convenient to untroduce the notion
\begin{eqnarray}
\label{sp3}
&& g^2\tilde{\Pi}_{(nl)}(p^2)=\lambda J_{(nl)}(\xi,b).
\end{eqnarray}

The dimensionless coupling constant can be defined as
\begin{eqnarray}
\label{sp4}
&& h_{(nl)}={1\over4\pi}{\xi \over[J_{(nl)}(\xi,b)]_b'}.
\end{eqnarray}

The numerical results are given in the table.

\begin{center}
\begin{tabular}{|c|c|c|c|c|c|c|c|} \hline
&&\multicolumn{6}{|c|}{$(nl)$}\\  \cline{3-8}
${m\over M}$ & $\lambda$ & \multicolumn{2}{|c|}{$(00)$}
& \multicolumn{2}{|c|}{$(10)$} & \multicolumn{2}{|c|}{$(01)$} \\
\cline{3-8}
&&{$\mu/M$}&{$h_{(00)}$}
&{$\mu/M$}&{$h_{(10)}$}&{$\mu/M$}&{$h_{(01)}$}\\
\hline
    &      &      &    &&&&   \\
.1  & 449  & 0.2  & 0.22 & 1.30 & .15 & 1.60 & .044   \\
.1  & 412  & 0.5  & 0.20 & 1.36 & .12 & 1.61 & .039  \\
.1  & 292  & 1.0  & 0.12 & 1.52 & .07 & 1.65 & .026  \\
    &        &      &    &&&&   \\
.3  & 129  & 0.2  & 0.71 & 1.55 & .43 & 1.53 & .140  \\
.3  & 120  & 0.5  & 0.64 & 1.59 & .38 & 1.54 & .130  \\
.3  & 92   & 1.0  & 0.41 & 1.72 & .24 & 1.59 & .085  \\
    &        &      &    &&&&   \\
.5  & 90   & 0.2  & 1.64 & 1.78 & .83 & 1.51 & .28  \\
.5  & 85   & 0.5  & 1.49 & 1.81 & .72 & 1.52 & .25  \\
.5  & 67   & 1.0  & 1.01 & 1.90 & .44 & 1.56 & .18  \\
    &        &      &    &&&&   \\
\hline
\end{tabular}
\end{center}

\subsection{Conclusion}

In conclusion the following points should be emphasized:
\begin{itemize}
\item the effective Lagrangian ${\cal L}_{eff}$ is Hermitian,
\item the effective interaction Lagrangian ${\cal I}_{int}$ does not
depend on the input coupling constant $g$,
\item the input coupling constant $g$ defines the spectrum of
bound states only,
\item the dimensionless coupling constant
$\lambda={1\over4\pi}\left({g\over M}\right)^2$ should be large,
\item the effective coupling constants $h_{(nl)}$ depend on the mass
of the corresponding bound state with the quantum number $(nl)$,
\item the effective coupling constants $h_{(nl)}$ are small, so that
the effective interaction Lagrangian ${\cal L}_{eff}$ can be computed
by the perturbation method,
\item in principle, the quadratic part of the total Lagrangian
should be diagonalized, i.e., the matrix $U_{QQ'}(p)$ should be found
for which
$$ \left[\delta_{QQ'}-g^2\tilde{\Pi}_{QQ'}(p)\right]=
\sum_{Q_1}U_{QQ_1}(p)\tilde{\Sigma}_{Q_1}(p)U_{Q_1Q'}^\top(p).$$
\end{itemize}

\subsection{Appendix}

The solution of (\ref{for7}) can be represented in the form of the following
functional integral (see Appendix and, for example, \cite{Dineykhan}):
\begin{eqnarray}
\label{app1}
&& S(x,y|\phi)={1\over\Box-M^2+g\phi(x)}\cdot\delta(x-y)\nonumber\\
&& ={1\over2}\int\limits_0^\infty d\alpha e^{-{\alpha\over2} M^2}
{\rm T}_\tau\exp\left\{{1\over2}\int\limits_0^\alpha d\tau
\left({\partial\over\partial x(\tau)}\right)^2
+{g\over2}\int\limits_0^\alpha d\tau \phi(x(\tau))\right\}\delta(x-y)
\nonumber\\
&& ={1\over2}\int\limits_0^\infty d\alpha e^{-{\alpha\over2} M^2}
\int D\nu\exp\left\{-{1\over2}\int\limits_0^\alpha d\tau\nu^2(\tau)\right\}
\nonumber\\
&& \cdot{\rm T}_\tau\left\{\int\limits_0^\alpha d\tau
\left(\nu(\tau){\partial\over\partial x(\tau)}\right)
+{g\over2}\int\limits_0^\alpha d\tau \phi(x(\tau))\right\}\delta(x-y)
\nonumber\\
&& ={1\over2}\int\limits_0^\infty d\alpha e^{-{\alpha\over2} M^2}
\cdot\delta\left(x-y+\int\limits_0^\alpha d\tau\nu(\tau)\right)\nonumber\\
&&\cdot \int D\nu\exp\left\{-{1\over2}\int\limits_0^\alpha d\tau\nu^2(\tau)+
{g\over2}\int\limits_0^\alpha
d\tau \phi\left(x+\int\limits_\tau^\alpha d\tau'\nu(\tau')\right)\right\}.
\nonumber
\end{eqnarray}
The function
$$ {\cal D}(x,\alpha;y,0\vert\phi)$$
$$={\rm T}_\tau\exp\left\{{1\over2}\int\limits_0^\alpha d\tau
\left({\partial\over\partial x(\tau)}\right)^2-{M^2\alpha\over2}
+{g\over2}\int\limits_0^\alpha d\tau \phi(x(\tau))\right\}\delta(x-y)$$
is the Green function of the equation
\begin{eqnarray}
&& -{\partial\over\partial\alpha}Y(x,\alpha)=
\left[-{1\over2}\left({\partial\over\partial x}\right)^2+{M^2\over2}-
{g\over2}\phi(x)\right]Y(x,\alpha),
\end{eqnarray}
i.e.
$$ Y(x,\alpha)=\int dy{\cal D}(x,\alpha;y,0\vert\phi)Y(y,0).$$
This equation can be considered as the Schr\"{o}dinger equation
where $\alpha$ plays the role of the imaginary time or "temperature",
$x\in{\rm R}^4$ and $\phi(x)$ is the Gaussian random potential
with the correlation function
$$ \langle\phi(x)\phi(y)\rangle_\phi=D_m(x-y).$$

The Green function $S(x,y|\phi)$ can be written in a more suitable
form. Let us introduce the variables
$$ \nu(\tau)=\nu_0+\mu(\tau),~~~~~~~\int\limits_0^\alpha d\tau\mu(\tau)=0$$
and integrate over $\nu_0$ using the $\delta$-function in the
representation (\ref{app1}). Introducing the variable
$$ \xi(\tau)=y\left(1-{\tau\over\alpha}\right)+x{\tau\over\alpha}
+\int\limits_\tau^\alpha d\tau'\mu(\tau')$$
one can get after some transformations
\begin{eqnarray}
&& S(x,y\vert\phi)=\int d\Sigma
\exp\left\{{g\over2}\int\limits_0^\alpha d\tau\phi
\left(\xi(\tau)\right)\right\},\nonumber\\
&& d\Sigma={d\alpha\over8\pi^2\alpha^2}e^{-\alpha M^2}D\xi
\exp\left\{-\int\limits_0^\alpha d\tau{\dot{\xi}^2(\tau)\over2}\right\},
\nonumber\\
&& \xi(0)=y,~~~~~~~~\xi(\alpha)=x.\nonumber
\end{eqnarray}

\end{document}